\documentclass[aps,preprint]{revtex4}
\usepackage[dvips]{graphicx}
\begin{document}
\draft
\title{\bf Investigation of nodal domains in a chaotic three-dimensional microwave
rough billiard with the translational symmetry}
\author{Nazar Savytskyy, Oleg Tymoshchuk, Oleh Hul, Szymon Bauch and Leszek Sirko}

\address{Institute of Physics, Polish Academy of Sciences, Aleja  Lotnik\'{o}w 32/46, 02-668 Warszawa, Poland}
\date{March 20, 2007}

\bigskip

\begin{abstract}

We show that using the concept of the two-dimensional level number
$N_{\bot}$ one can experimentally study of the nodal domains in a
three-dimensional (3D) microwave chaotic rough billiard with the
translational symmetry. Nodal domains are regions where a wave
function has a definite sign. We found the dependence of the
number of nodal domains $\aleph_{N_{\bot}}$ lying on the
cross-sectional planes of the cavity on the two-dimensional level
number $N_{\bot}$. We demonstrate that in the limit $N_{\bot}
\rightarrow \infty $ the least squares fit of the experimental
data reveals the asymptotic ratio $\aleph_{N_{\bot}}/N_{\bot}
\simeq 0.059 \pm 0.029$ that is close to the theoretical
prediction $\aleph_{N_{\bot}}/N_{\bot} \simeq 0.062$. This result
is in good agreement with the predictions of percolation theory.

\end{abstract}


\bigskip
\maketitle

\smallskip

In this paper we show that measuring the  distributions of the
electric field of TM modes of a 3D chaotic rough cavity with the
translational symmetry one can find the dependence of the number
of nodal domains $\aleph_{N_{\bot}}$ lying on the cross-sectional
planes of the cavity on the two-dimensional level number
$N_{\bot}$. The translational symmetry means that the
cross-section of the billiard is invariant under translation along
$z$ direction.

In the seminal papers Blum {\it et al.} \cite{Blum2002} and
Bogomolny and Schmit \cite{Bogomolny2002} showed that the
distributions of the number of nodal domains in two-dimensional
(2D) systems can be used to distinguish between the systems with
integrable and chaotic underlying classical dynamics. The
theoretical findings have been tested in a series of experiments
with chaotic microwave 2D rough billiards
\cite{Savytskyy2004,Hul2005,Hul2006}.

Due to severe experimental problems there are very few
experimental studies devoted to 3D chaotic microwave cavities
\cite{Sirko1995,Alt1997,Dorr1998,Eckhardt1999,Dembowski2002}. In a
pioneering experiment Deus {\it et al.} \cite{Sirko1995} have been
measured eigenfrequencies of the 3D chaotic (irregular) microwave
cavity in order to confirm that their distribution displays
behavior characteristic for classically chaotic quantum systems,
viz., the Wigner distribution.  In other important experiments the
periodic orbits \cite{Alt1997}, the distributions and the
correlation function  of the frequency shifts caused by the
external perturbation \cite{Dorr1998,Eckhardt1999} and a trace
formula for chaotic 3D cavities \cite{Dembowski2002} have been
respectively studied. Quite recently the  spatial correlation
functions of the 3D experimental microwave chaotic rough billiard
with the translational symmetry have been studied by Tymoshchuk et
al \cite{Tymoshchuk2007}. Three-dimensional chaotic cavities and
properties of random electromagnetic vector field have been also
studied in several theoretical papers
\cite{Primack2000,Prosen1997,Arnaut2006}.

The important feature of 3D cavities with the translational
symmetry is connected with the fact that their modes can be
classified into transverse electric (TE) and transverse magnetic
(TM). Although, there is no analogy between quantum billiards and
electromagnetic cavities in three dimensions, the TM modes are
especially important because they allow for the simulation of 2D
quantum billiards on cross-sectional planes of 3D cavities.

\begin{figure}[!]
\begin{center}
\rotatebox{270} {\includegraphics[width=0.5\textwidth,
height=0.6\textheight, keepaspectratio]{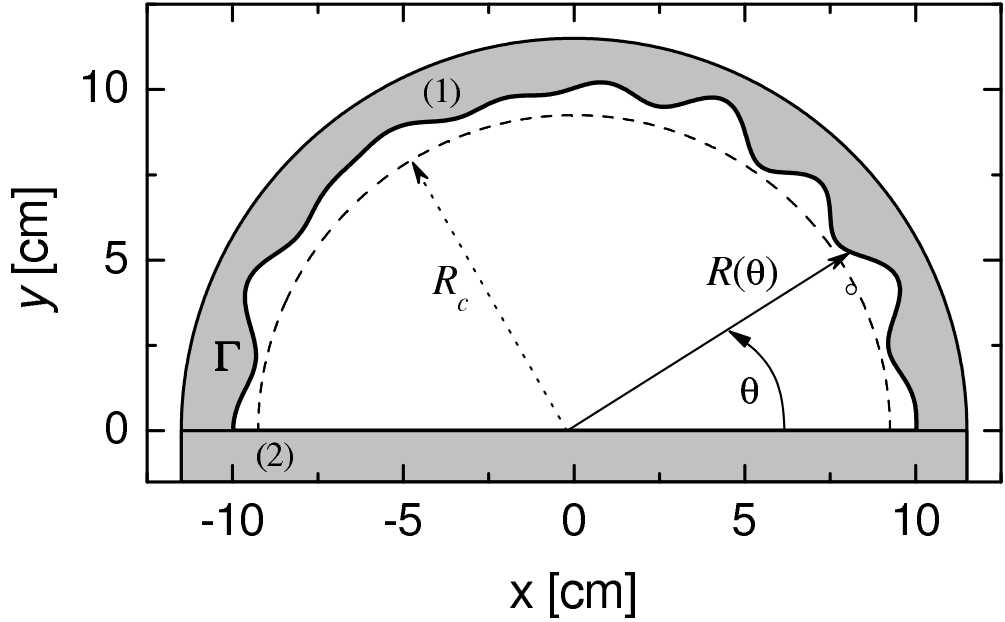}} \caption{Sketch of
the chaotic half-circular 3D microwave rough billiard in the $xy$
plane. Dimensions are given in cm. The cavity sidewalls are marked
by 1 and 2 (see text). Squared wave functions
$|\psi_{N,p}(R_c,\theta )|^2$  were evaluated on a half-circle of
fixed radius $R_c=9.25$ cm. Billiard's rough boundary $\Gamma $ is
marked with the bold line. The white circle centered at $x=8.12$
cm and $y=4.13$ cm marks the position of the hole drilled in the
upper wall of the cavity. The hole was used to introduce the
perturber inside the cavity in order to measure the $z$-component
of the electric field distributions $E_{N,p}({\bf
x})$.}\label{Fig1}
\end{center}
\end{figure}

In the experiment we used 3D cavity with the translational
symmetry in the shape of a rough half-circle (Fig.~\ref{Fig1})
with the height $h=60$ mm. The cavity was made of polished
aluminium. The upper and bottom walls of the cavity were attached
to the sidewalls with 48 screws in order to make good electrical
contact.

Assuming that the direction of the translational symmetry of the
cavity is along the $z$-axis the boundary conditions at $z=0$ and
$z=h$ demand that the $z$ dependence of the $z$-component of the
electric and magnetic fields $E_{N,p}({\bf x})$ and $B_{N,p}({\bf
x})$  of TM modes be in the form $E_{N,p}({\bf x})\equiv
E_{N,p}(x,y,z)=A_{N,p} \psi_{N,p}(x,y) f_{p}(z)$, where
$f_{p}(z)=\cos(p\pi z/h)$, $p=0,1,2\ldots$, $A_{N,p}$ is the
normalization constant and $B_{N,p}({\bf x})=0$. The dependence of
$E_{N,p}({\bf x})$ on the plane cross section coordinates we
denote by the amplitude $\psi_{N,p}(x,y)\equiv E_{N,p}(x,y)$.
Then, the amplitude $\psi_{N,p}(x,y)$ satisfies the Helmholtz
equation
$$
(\bigtriangleup _{\bot} +k_{N,p}^2)\psi_{N,p}(x,y)=0, \eqno(1)
$$
where $\bigtriangleup _{\bot}$ is two-dimensional Laplacian
operator and $k_{N,p}= (k_N^2-(p\pi/h)^2)^{1/2}$ is the effective
wave vector. The wave vector $k_N=2\pi \nu_N/c$, where $\nu_N$ is
the resonance frequency of the level $N$ and $c$ is the speed of
light in the vacuum. One can easily see that the equation (1) is
equivalent to the Schr\"odinger equation (in units $\hbar =1$)
describing a particle of mass $m=1/2$ with the kinetic energy
$k_N^2$ in an external potential $V=(p\pi/h)^2$ \cite{Kim2005}.
Therefore, 3D microwave cavities can be effectively used  beyond
the standard 2D frequency limit (the case $p=0$) \cite{Hans} in
simulation of quantum systems. The amplitude $\psi_{N,p}(x,y)$
fulfills Dirichlet boundary conditions on the sidewalls of the
billiard and therefore, throughout the text it is also called the
wave functions $\psi_N(x,y)$. It is worth noting that the full
electric field $E_{N,p}({\bf x})$  satisfies  Neumann boundary
conditions at the top and the bottom of the cavity.

The measurements of $E_{N,p}({\bf x})$ of a 3D microwave cavity
allowed us  to test experimentally an important finding of the
papers by Blum {\it et al.} \cite{Blum2002} and Bogomolny and
Schmit \cite{Bogomolny2002} which  connects  the number of nodal
domains of 2D billiards with the level number $N$. We will show
that for the 3D cavities with the translational symmetry the
number of nodal domains $\aleph_{N_{\bot}}$ lying on the
cross-sectional planes of the cavity is connected  with the
two-dimensional level number $N_{\bot}$. The condition
$E_{N,p}({\bf x})|_{z=const}=0$ on the cross-sectional planes of
the cavity determines a set of nodal lines which separate regions
(nodal domains) with opposite signs of the  electric field
distribution $E_{N,p}({\bf x})|_{z=const}$.

The value of the level number $N$ of the 3D cavity was evaluated
from the Balian--Bloch formula \cite{Balian}.
$$
N(k) = \frac{1}{3\pi ^2}Vk^3 -\frac{2}{3\pi^
2}\int_S\frac{d\sigma_{\omega}}{R_{\omega}}k,\eqno(2)
$$
where k is the wave vector, $V=(9.43 \pm 0.01)\cdot 10^{-4}$ m$^3$
is the volume of the cavity and
$\int_S\frac{d\sigma_{\omega}}{R_{\omega}}=0.932$ m $\pm 0.005$ m
is the surface curvature averaged over the surface of the cavity.
We used this formula because of the relatively low quality factor
of the cavity ($Q \simeq 4000$) some resonances overlapped.

The two-dimensional level number $N_{\bot}$ is defined by the
standard Weyl--Bloch formula  $N_{\bot} =
\frac{A}{4\pi}k_{N,p}^2-\frac{P}{4\pi}k_{N,p}$, where $A=(1.572
\pm 0.002)\cdot 10^{-2}$ m$^{2}$ and $P=0.537$ m  $\pm 0.001$ m
are the cross-sectional plane area of the cavity and its
perimeter, respectively.

The cavity sidewalls consist of two segments (see
Fig.~\ref{Fig1}). The rough segment 1 is described on the
cross-sectional planes by the radius function
$R(\theta)=R_{0}+\sum_{m=2}^{M}{a_{m}\sin(m\theta+\phi_{m})}$,
where  the mean radius $R_0$=10.0 cm, $M=20$, $a_{m}$ and
$\phi_{m}$ are uniformly distributed on [0.084,0.091] cm and
[0,2$\pi$], respectively, and $0\leq\theta<{\pi}$. (For
convenience, the polar coordinates $r$ and $\theta$ are used
instead of the Cartesian ones $x$ and $y$.) It is worth noting
that following our earlier experience
\cite{Hlushchuk01b,Hlushchuk01} we decided to use a rough
desymmetrized half-circular cavity instead of a rough circular
cavity, because the first one  lowers the number of nearly
degenerated eigenfrequencies. Additionally, a half-circular
geometry of the cavity was suitable in the procedure of  accurate
measurements of the electric field distributions inside the
billiard.

The roughness of a billiard on the cross-sectional planes can be
characterized by the function $k(\theta)=(dR/d\theta)/R_0$. For
our  microwave billiard we have the angle average $\tilde k=(\left
<k^{2}(\theta)\right >_{\theta})^{1/2}\simeq   0.400$. The value
of $\tilde k $ is much above the chaos border
$k_c=M^{-5/2}=0.00056$ \cite{Frahm97} which indicates that in such
a billiard the classical dynamics is diffusive in orbital momentum
due to collisions with the rough boundary.

The other properties of the billiard \cite{Frahm} are also
determined by the roughness parameter $\tilde k $. The amplitudes
$\psi_{N,p}(r,\theta)$ are localized for the two-dimensional level
number $N_{\bot} < N_e = 1/128 \tilde k^4$. Because of a large
value of the roughness parameter $\tilde k $ the localization
border lies very low, $N_e \simeq 1$. The border of Breit-Wigner
regime is $N_W = M^2/48\tilde k^2 \simeq 52$. It means that
between $N_e < N_{\bot} < N_W$ Wigner ergodicity \cite{Frahm}
ought to be observed and for $N_{\bot} > N_W$ Shnirelman
ergodicity should emerge.

In order to measure the amplitudes $\psi_{N,p}(r,\theta)$ of the
3D electric field distributions we used an effective  method
described in \cite{Savytskyy2003}. It is based on the perturbation
technique and preparation of the ``trial functions". In this
method the amplitudes $\psi_N(r,\theta )$ (electric field
distribution $E_N(r, \theta )$ inside the cavity) are determined
from the form  of electric field $E_{N,p}(R_c, \theta )$ evaluated
on a half-circle of  fixed radius $R_c$ (see Fig. \ref{Fig1}). The
first step in evaluation of $E_{N,p}(R_c, \theta )$ is measurement
of $|E_{N,p}(R_c, \theta )|^2$. The perturbation technique
developed in \cite{Slater52} and used successfully in
\cite{Slater52,Sridhar91,Richter00,Anlage98} was implemented for
this purpose. In this method a small perturber is introduced
inside the cavity to alter its resonant frequency.

The perturber (4.0 mm in length and 0.3 mm in diameter, oriented
in $z$-direction) was moved by the stepper motor via the Kevlar
line hidden in the groove (0.4 mm wide, 1.0 mm deep) made in the
cavity's bottom wall along the half-circle $R_c$. Before closing
the cavity we carefully inspected whether the pin moves smoothly,
oriented in vertical position. Using such a perturber we had no
positive frequency shifts that would exceed the uncertainty of
frequency shift measurements (15 kHz).

In order to determine the dependence of the electric field
distributions $E_{N,p}({\bf x})$ on the $z$ coordinate and to
estimate the wave vector $k_3=p\pi/h$ we measured the electric
field inside the 3D cavity along the $z$-axis. The perturber (4.5
mm in length and 0.3 mm in diameter) was attached to the Kevlar
line and moved by the stepper motor. It entered and exited the
cavity by small holes (0.4 mm) drilled in the upper and the bottom
walls of the cavity. The both holes were located at the position:
$r=9.11$ cm, $\theta=0.47$ radians.

To eliminate the variation of resonant frequencies connected with
the thermal expansion of the aluminium cavity  the temperature of
the cavity was stabilized with the accuracy of 0.05 $\deg $.

Using a field perturbation technique we were able to measure
squared wave functions $|\psi_{N,p}(R_c,\theta )|^2$  for 80 TM
modes within the region $2\leq N \leq 489$. The range of
corresponding eigenfrequencies was from $\nu_{2} \simeq 2.47$ GHz
to $\nu_{489} \simeq 11.99$ GHz. The measurements were performed
at 0.36 mm steps along a half-circle with fixed radius $R_c=9.25$
cm. This step was small enough to reveal in details the space
structure of high-lying levels.

\begin{figure}[!]
\begin{center}
\rotatebox{0} {\includegraphics[width=0.5\textwidth,
height=0.8\textheight, keepaspectratio]{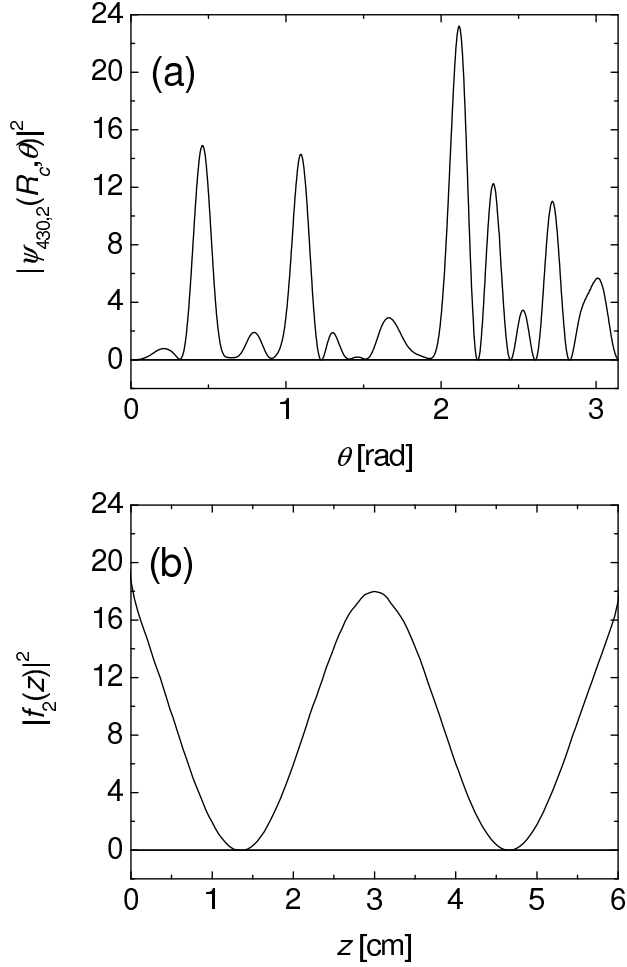}} \caption{Panel
(a): Squared wave function $|\psi_{430,2}(R_c,\theta )|^2$ (in
arbitrary units) measured on a half-circle with radius $R_c=9.25$
cm ($\nu_{430} \simeq 11.50$ GHz). Panel (b): Squared
$z$-component of the electric field distribution $|f_{2}(z)|^2$
measured at $r=9.11$ cm and $\theta=0.47$ radians.}\label{Fig2}
\end{center}
\end{figure}

In Fig. \ref{Fig2} (a) and Fig. \ref{Fig2} (b) we show the
examples of the squared amplitude $|\psi_{N,p} (R_c, \theta )|^2$
and the squared $z$-component of the field, respectively,
evaluated for the level number $N=430$.

The perturbation method used in our measurements allows us to
extract information about the modulus of the wave function
amplitude $|\psi_{N,p}(R_c, \theta )|$ at any given point of the
cross-sectional plane $z=0$  but it doesn't allow to determine the
sign of $\psi_{N,p}(R_c, \theta )$. In order to obtain information
about the sign of $\psi_{N,p}(R_c, \theta )$ we used the method of
the ``trial wave function" precisely described in
 \cite{Savytskyy2003,Savytskyy2004,Hul2005}.

The amplitudes $\psi_{N,p}(r, \theta )$ of the electric field
distributions of a rough half-circular 3D billiard may be expanded
in terms of circular waves (here only odd states in expansion are
considered)
$$
\psi_{N,p}(r, \theta ) = \sum_{s=1}^L a_s
J_{s}(k_{N,p}r)\sin(s\theta ), \eqno(5)
$$
where $J_{s}$ is the Bessel function of order $s$.

In Eq. (5) the number of basis functions is limited to $L=k_{N,p}
r_{max} = l_{N}^{max}$, where $r_{max}=10.64$ cm is the maximum
radius of the cavity. $l_N^{max} = k_{N,p} r_{max}$ is a
semiclassical estimate for the maximum possible angular momentum
for a given $k_N$. Circular waves with angular momentum $s > L$
correspond to evanescent waves and can be neglected. Coefficients
$a_s$ may be extracted from the ``trial wave function"
$\psi_{N,p}(R_c, \theta )$ via
$$
a_s=[\frac{\pi }{2}J_{s}(k_{N,p}R_c)]^{-1}\int_0^{\pi}\psi_{N,p}
(R_c,\theta )\sin(s\theta )d\theta . \eqno(6)
$$
Due to experimental uncertainties and the finite step size in the
measurements of $|\psi_{N,p}(R_c, \theta )|^2$ the wave functions
$\psi_{N,p}(r, \theta )$ are not exactly zero at the boundary
$\Gamma $.  As the quantitative measure of the sign assignment
quality we chose  the integral $I = \gamma \int_{\Gamma
}|\psi_{N,p}(r,\theta )|^2dl $ calculated along the billiard's
rough boundary $\Gamma $, where $\gamma $ is length of $\Gamma $.
For correctly reconstructed wave functions the integral $I$ was
several times smaller than in the case of not correctly
reconstructed ones.

It is worth noting that since the pin is attached to the line it
cannot be stuck. However, one may assume that during the movement
the pin may be accidentally, from time to time, slightly slanted,
adding small "a noise-like component" to the measured electric
field. The formula (5) shows that each wave function is expanded
in terms of $L$ circular waves  which filters out noise-like
higher frequency Fourier components from the reconstructed wave
function. The same filtering removes out the influence of the
experimental uncertainties of frequency shifts on the
reconstructed wave functions.

\begin{figure}[!]
\begin{center}
\rotatebox{270} {\includegraphics[width=0.5\textwidth,
height=0.6\textheight, keepaspectratio]{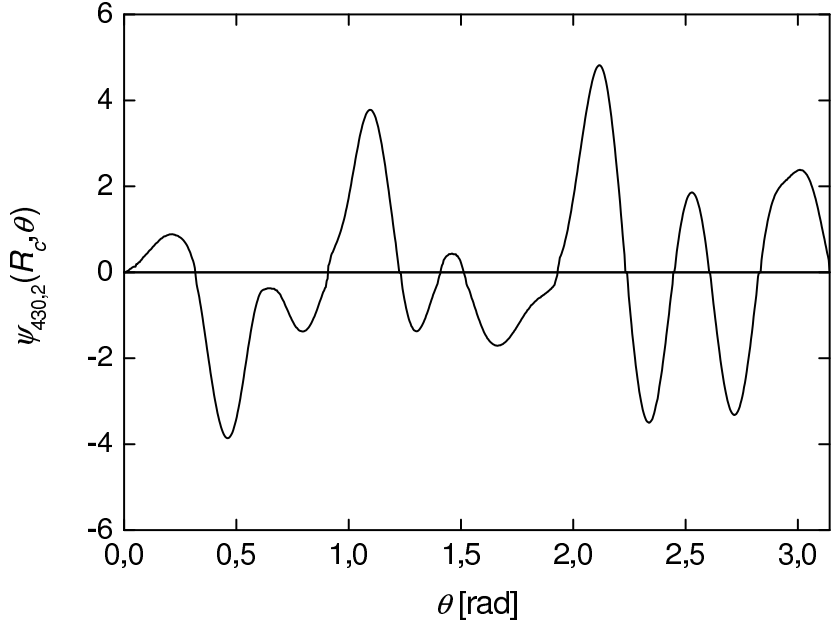}} \caption{The
``trial wave function" $\psi_{430,2}(R_c,\theta )$ (in arbitrary
units) with the correctly assigned  signs, which was used in the
reconstruction of the wave function $E_{430,2}(r, \theta,z )$ of
the billiard (see Fig. \ref{Fig4}). }\label{Fig3}
\end{center}
\end{figure}

\begin{figure}[!]
\begin{center}
\rotatebox{0} {\includegraphics[width=0.5\textwidth,
height=0.8\textheight, keepaspectratio]{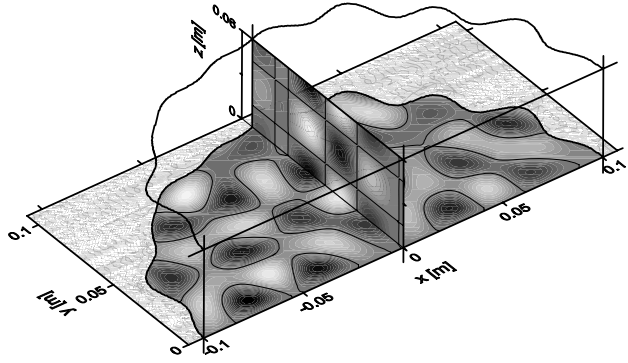}} \caption{ The
reconstructed wave function $\psi_{430,2}(r,\theta) $ of the
chaotic half-circular microwave rough billiard. The amplitudes
have been converted into a grey scale with white corresponding to
large positive and black corresponding to large negative values,
respectively. The structure of the nodal lines are shown by the
bold full lines. Dimensions of the billiard are given in cm. In
the figure the $z$ dependence of the electric field distribution $
E_{430,2}(r,\theta,z )|_{x=0} \propto
\psi_{430,2}(r,\theta)|_{x=0} f_{2}(z)$ is also shown.
}\label{Fig4}
\end{center}
\end{figure}

In Fig. \ref{Fig3} we show the ``trial wave function"
$\psi_{430,2} (R_c, \theta )$ with the correctly assigned  signs,
which was used in the  reconstruction of the wave function
$\psi_{430,2}(r, \theta )$ of the billiard (see Fig. \ref{Fig4}).
In Fig. \ref{Fig4} different nodal domains are separated by the
bold full lines.

Using the method of the ``trial wave function" we were able to
reconstruct 75 experimental wave functions $\psi_{N,p}(r, \theta
)$, which belonged to TM modes of the rough half-circular 3D
billiard with the level number $N$ between 2 and 489.  The
remaining wave functions belonging to TM modes, from the range
$N=2-489$,  were not reconstructed because of near-degeneration of
the neighboring eigenfrequencies or due to the problems with the
measurements of $|\psi_{N,p}(R_c, \theta )|^2$ along a half-circle
coinciding for its significant part with one of the nodal lines of
$\psi_{N,p}(r, \theta )$.

\begin{figure}[!]
\begin{center}
\rotatebox{0} {\includegraphics[width=0.5\textwidth,
height=0.8\textheight, keepaspectratio]{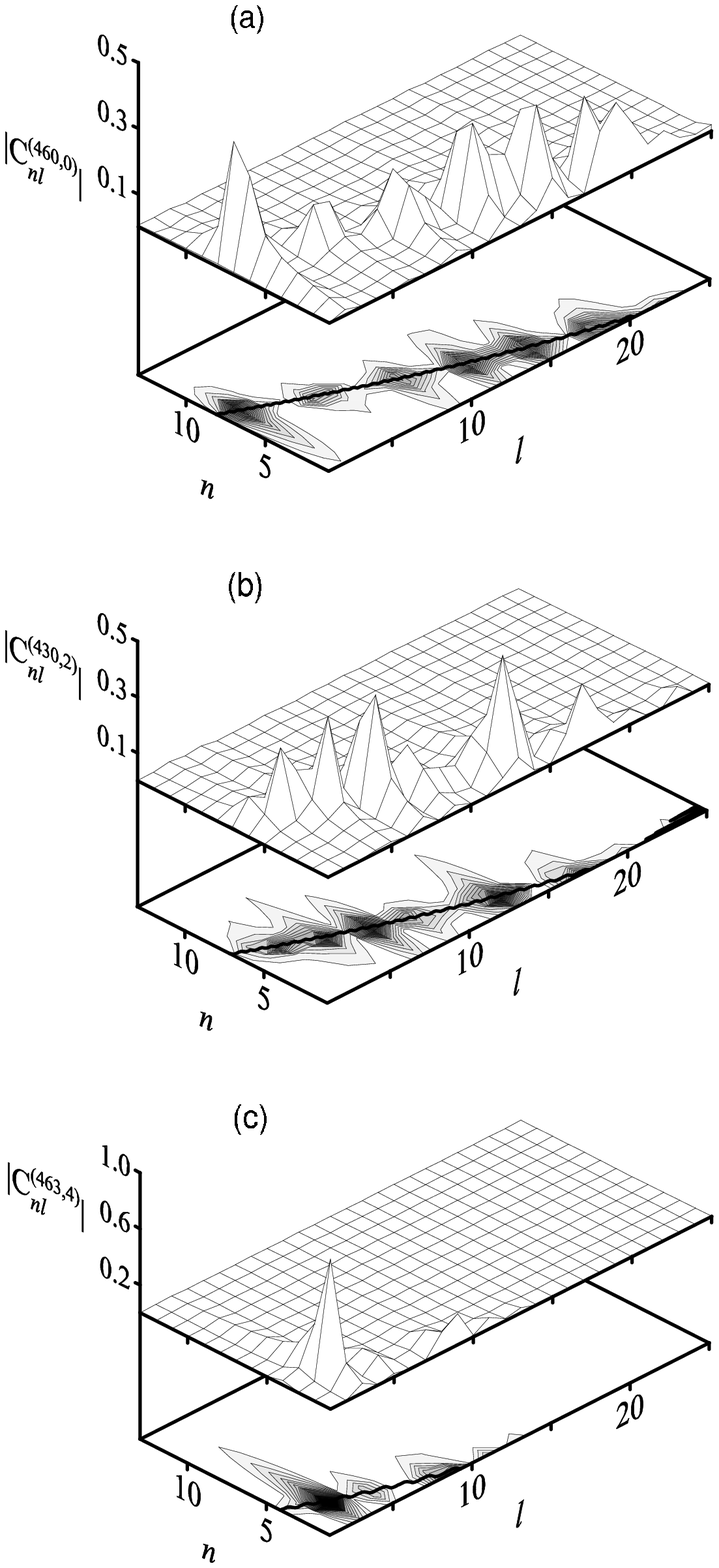}} \caption{Structure
of the energy surface of the wave functions lying close to the
boarder of the regimes of Breit-Wigner and Shnirelman ergodicity
($N_{W} = 52$), panels (a) and (b), and for the low wave function
in the regime of Breit-Wigner ergodicity, panel (c). Panel (a):
The moduli of amplitudes $|C^{(460,0)}_{nl}|$ for the wave
function $\psi_{460,0}(r,\theta)$, $N_{\bot} = 65$, lying in the
regime of Shnirelman ergodicity. Panel (b): The moduli of
amplitudes $|C^{(430,2)}_{nl}|$ for the wave function
$\psi_{430,2}(r,\theta)$, $N_{\bot} = 50$, in the regime of
Breit-Wigner ergodicity . Panel (c): The moduli of amplitudes
$|C^{(463,4)}_{nl}|$ for the wave function
$\psi_{463,4}(r,\theta)$ lying in the regime of Breit-Wigner
ergodicity close to the localization boarder. Full lines show the
semiclassical estimation of the energy surface (see
text).}\label{Fig5}
\end{center}
\end{figure}

The borders of Breit-Wigner and Shnirelman ergodicities are not
sharp. Therefore, to check ergodicity of the billiard's wave
functions $\psi_{N,p}(r, \theta )$, especially close to the
borders, one should use some additional measures  such as  e.g.,
calculation of the structures of their energy surfaces
\cite{Frahm97}.  For this reason we extracted wave function
amplitudes $C^{(N,p)}_{nl}=\left< n,l|N,p \right>$ in the basis
$n, l$ of a half-circular billiard with radius $r_{max}$, where $
n=1,2,3 \ldots$ enumerates the zeros of the Bessel functions and $
l=1,2,3 \ldots$ is the angular quantum number. The moduli of
amplitudes $|C^{(N,p)}_{nl}|$  and their projections into the
energy surface for the experimental wave functions
$\psi_{460,0}(r, \theta )$, $\psi_{430,2}(r, \theta )$ and
$\psi_{463,4}(r, \theta )$ are shown in Fig. \ref{Fig5}(a-c). As
expected, on the border of the regimes of Breit-Wigner and
Shnirelman ergodicity the wave functions $\psi_{460,0}(r, \theta
)$ ($N_{\bot} = 65$)  and $\psi_{430,2}(r, \theta )$ ($N_{\bot} =
50$) are extended homogeneously over the whole energy surface
\cite{Hlushchuk01}. The wave function $\psi_{463,4}(r, \theta )$,
$N_{\bot} = 16$, which lies closer to the localization boarder, is
also extended along the energy surface, however it displays the
tendency to localization in $n, l$ basis (see Fig. \ref{Fig5}(c)).
The full lines on the projection planes in Fig. \ref{Fig5}(a-c)
mark the energy surface of a half-circular billiard
$H(n,l)=k^2_{N,p}$ estimated from the semiclassical formula
\cite{Hlushchuk01b}: $\sqrt{(l^{max}_{N})^2 - l^2}
-l\arctan(l^{-1}\sqrt{(l^{max}_{N})^2 - l^2}) + \pi/4 = \pi n $.
It is clearly visible that the peaks $|C^{(N,p)}_{nl}|$ are spread
almost perfectly along the lines marking the energy surface.

\begin{figure}[!]
\begin{center}
\rotatebox{270} {\includegraphics[width=0.5\textwidth,
height=0.6\textheight, keepaspectratio]{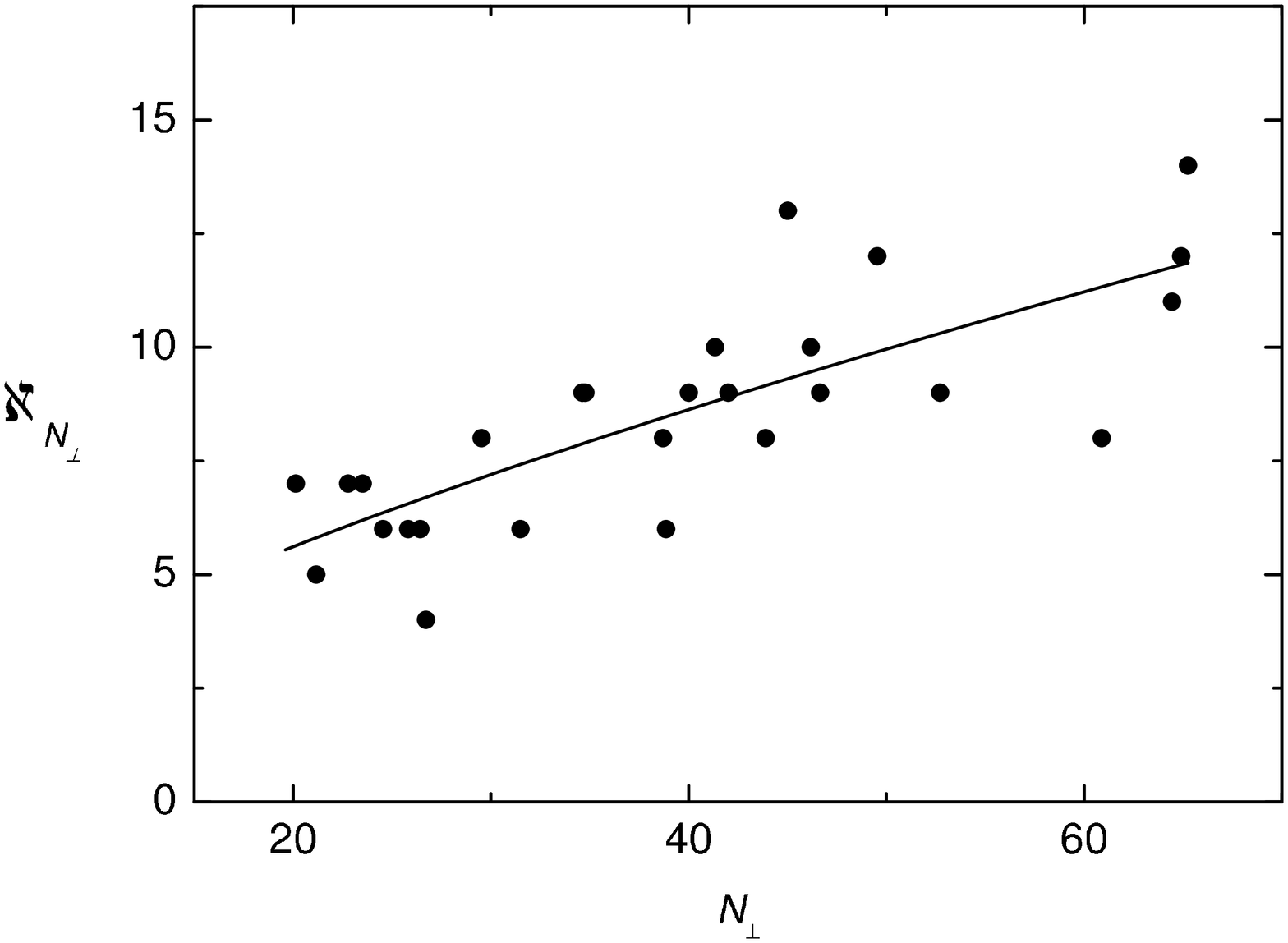}} \caption{The
number of nodal domains $\aleph_{N_{\bot}}$ (full circles) on the
cross-section planes of the chaotic half-circular 3D microwave
rough billiard. Full line shows the least squares fit
$\aleph_{N_{\bot}} = a_1N_{\bot} +b_1\sqrt{N_{\bot}}$ to the
experimental data (see text), where $a_1=0.059 \pm 0.029$,
$b_1=0.991 \pm 0.190$. The prediction of the theory of Bogomolny
and Schmit \cite{Bogomolny2002} $a_1=0.062$. }\label{Fig6}
\end{center}
\end{figure}

The number of nodal domains $\aleph_{N_{\bot}}$ on the
cross-sectional plane $z=0$ vs. the level number $N_{\bot}$ in the
chaotic 3D microwave rough billiard is plotted in Fig. \ref{Fig6}.
The full line in Fig. \ref{Fig6} shows the least squares fit
$\aleph_{N_{\bot}} = a_1N_{\bot} +b_1\sqrt{N_{\bot}}$ of the
experimental data, where $a_1=0.059 \pm 0.029$, $b_1=0.991 \pm
0.190$. The coefficient $a_1=0.059 \pm 0.029$ coincides with the
prediction of the percolation model of Bogomolny and Schmit
\cite{Bogomolny2002} $\aleph_{N_{\bot}}/N_{\bot} \simeq 0.062$
within the error limits. The relatively large uncertainty of the
coefficient $a_1$ is connected with the fact that in the least
squares fit procedure we used only 27 higher states with
$N_{\bot}> 20$. The states with lower $N_{\bot}$  were not taken
into account because  they were not fully chaotic (see Fig.
\ref{Fig5}(c)). The second term in the least squares fit
corresponds to a contribution of boundary domains, i.e. domains,
which include the billiard boundary. Numerical calculations of
Blum {\it et al.} \cite{Blum2002} performed for the Sinai and
stadium billiards showed that the number of boundary domains
scales as the number of the boundary intersections, that is as
$\sqrt{N_{\bot}}$. Our results clearly suggest that in the rough
billiard, at the level numbers $20 < N_{\bot} \leq 65$, the
boundary domains also significantly influence the scaling of the
number of nodal domains $\aleph_{N_{\bot}}$, leading to the
departure from the predicted scaling $\aleph_{N_{\bot}} \sim
N_{\bot}$.

In summary, we measured the wave functions of the chaotic 3D rough
microwave billiard with the translational symmetry up to the level
number $N=489$.  We showed that for the two-dimensional level
numbers $20 < N_{\bot} \leq 65$ the scaling of the number of nodal
domains $\aleph_{N_{\bot}}$  significantly departures from the
predicted scaling $\aleph_{N_{\bot}} \sim N_{\bot}$, which
suggests that the boundary domains influence the scaling
\cite{Bogomolny2002}. In the limit $N_{\bot} \rightarrow \infty $
the least squares fit of the experimental data yields the
asymptotic number of nodal domains $\aleph_{N_{\bot}}/N_{\bot}
\simeq a_1=0.059 \pm 0.029$ that is close to the theoretical
prediction $\aleph_{N_{\bot}}/N_{\bot} \simeq 0.062$. Finally, our
results show that 3D microwave cavities with the translational
symmetry can be effectively used beyond the standard 2D frequency
limit in simulation of quantum systems.

Acknowledgments.  This work was partially supported by the
Ministry of Education and Science grant No. N202 099 31/0746.

\end{document}